\input phyzzx
\Pubnum={\vbox{\hbox{NSF-ITP-96-07}\hbox{hep-th/9602071}}}
\date={}
\pubtype={}
\def\a{\alpha}
\def\b{\beta}
\def\g{\gamma}
\def\d{\delta}
\def\e{\epsilon}

\def\th{\theta}

\def\k{\kappa}

\def\n{\nu}

\def\s{\sigma}
\def\t{\tau}

\def\G{\Gamma}

\def\T{\Theta}

\def\S{\Sigma}

\def\o{\over}

\def\np{{\it Nucl. Phys. }}
\def\pl{{\it Phys. Lett. }}

\def\mpl{{\it Mod. Phys. Lett. }}
\def\prl{{\it Phys. Rev. Lett. }}

\def\no{\noindent}

\def\IR{\relax{\rm I\kern-.18em R}}
\font\cmss=cmss10 \font\cmsss=cmss10 at 7pt
\def\IZ{\relax\ifmmode\mathchoice
{\hbox{\cmss Z\kern-.4em Z}}{\hbox{\cmss Z\kern-.4em Z}}
{\lower.9pt\hbox{\cmsss Z\kern-.4em Z}}
{\lower1.2pt\hbox{\cmsss Z\kern-.4em Z}}\else{\cmss Z\kern-.4em Z}\fi}
\def\IN{\relax{\rm I\kern-.18em N}}
\def\IC{\relax{\rm I\kern-.18em C}}

\def\pa{\partial}
\def\tria{\triangleright}
\def\nsl{\nabla \!\!\!\! /}

\def\nsl{n \!\!\! /}
\def\tsl{t\!\! / }

\def\Tbar{\bar \Theta} 
\def\XD{X^I_{\scriptscriptstyle {\cal D}}}
\def\XN{X^I_{\scriptscriptstyle {\cal N}}}
\def\SSD{\scriptscriptstyle {\cal D}}
\def\SSN{\scriptscriptstyle {\cal N}}

\titlepage
\title{BOUNDARIES IN {\cal M}-THEORY}
\author{Katrin Becker\foot{beckerk@itp.ucsb.edu}}
\address{Institute for Theoretical Physics\break University of California
\break Santa Barbara, CA 93106-4030}
\author{Melanie Becker\foot{mbecker@denali.physics.ucsb.edu}}
\address{Department of Physics\break University of California \break
Santa Barbara, CA 93106-9530}
\abstract{We formulate boundary conditions for an open membrane 
that ends 
on the fivebrane of 
{\cal M}-theory. We show that the dynamics of the eleven-dimensional
fivebrane can be obtained from the quantization of a 
``small membrane'' that
is confined to a single fivebrane and which moves with 
the speed of light.
This shows that the eleven-dimensional fivebrane has an
interpretation as a $D$-brane of an open supermembrane as has 
recently been proposed by Strominger and Townsend.
We briefly discuss the boundary dynamics of an infinitely extended
planar membrane that is stretched between two parallel fivebranes.}

\endpage
\chapter{Introduction}
\no {\cal M}-theory is the ``magic'' quantum theory in eleven
dimensions from which properties of string theory, 
such as duality symmetries,
can be understood in a unified framework.
The low energy limit of this theory, eleven dimensional 
supergravity, contains
membranes and fivebranes. Properties of {\cal M}-theory 
can be analyzed, for example,  
by considering compactifications of these extended objects 
to lower dimensions
\REF\stdy{E.~Witten, ``String Theory Dynamics in Various Dimensions'', 
\np {\bf B443} (1995) 85, hep-th/9503124.}
\REF\bbs{K.~Becker, M.~Becker and A.~Strominger, ``Fivebranes, Membranes 
and Nonperturbative String Theory'', \np {\bf B456} (1995) 130, 
hep-th/9507158.}
\REF\Sch{J.~H.~Schwarz, ``An $SL(2,Z)$ Multiplet of 
Type IIB Superstrings'', 
\pl {\bf B360} (1995) 13, hep-th/9508143; ``The Power 
of $M$-Theory'', \pl {\bf B367} (1996) 97, hep-th/9510086; 
``$M$-Theory Extensions 
of $T$-Duality'', 
CALT-68-2034, hep-th/9601077.}
\REF\howi{P.~Ho\v rava and E.~Witten, ``Heterotic and Type 
I String Dynamics
from Eleven Dimension'', preprint IASSNS-HEP-95-86, 
PUPT-1571, hep-th/ 9510209.}
\REF\Hull{C.~M.~Hull, ``String Dynamics at Strong Coupling'', 
preprint QMW-95-50, hep-th/9512181.}
\REF\Mk{K.~Dasgupta and S.~Mukhi, ``Orbifolds of $M$-Theory'', preprint 
TIFR-TH-95-64, hep-th/9512196. }
\REF\Sen{A.~Sen, ``$T$-Duality of $p$-Branes'', preprint MRI-PHY-28-95, 
hep-th/9512203.}
\REF\Wit{E.~Witten, ``Five-Branes and $M$-Theory on an Orbifold'', 
preprint IASSNS-HEP-96-01, hep-th/9512219.}
\foot{See [\stdy,\bbs,\Sch ,\howi,\Hull,\Mk,\Sen,\Wit] and 
references therein.}.
 
All the ten-dimensional type IIA
$p$-branes can be related to either the membrane or 
the fivebrane 
\REF\mfi{P.~K.~Townsend, ``The Eleven-Dimensional Supermembrane 
Revisited'', \pl {\bf B350} (1995) 184, hep-th/9501068.}
[\mfi] of eleven dimensional supergravity. 
Polchinski
\REF\polch{J.~Polchinski, ``Dirichlet-Branes and Ramond-Ramond Charges'', 
\prl {\bf 75} (1995) 4724, hep-th/9510017.}
[\polch]
has shown that ten-dimensional Ramond-Ramond (RR) $p$-branes can be 
described as $D$-branes\foot{A recent review article is  
\REF\pcj{J.~Polchinski, S.~Chaudhuri and C.~V.~Johnson, ``Notes on 
$D$-Branes'', preprint NSF-ITP-96-003, hep-th/9602052.}[\pcj].}, 
that are hyperplanes on which an open 
string is allowed to end 
\REF\dlp{J.~Dai, R.~G.~Leigh and J.~Polchinski, ``New Connections 
Between String Theories'', \mpl {\bf A4} (1989) 2073.} 
[\dlp]. 
The appearance of Dirichlet boundary conditions can 
be naturally understood as a consequence 
of $T$-duality [\dlp]. 
String duality therefore teaches us that we have to incorporate
Dirichlet boundary conditions in open string theory in order to have
a consistent theory. Moreover, as explained in [\dlp], a $D$-brane is a 
dynamical object. Massless open string excitations that propagate on
the $D$-brane have an interpretation as collective coordinates for the 
transverse fluctuations of the $D$-brane. 
The recent work of Douglas
\REF\doug{M.~Douglas, ``Branes within Branes'', preprint 
RU-95-92, hep-th/9512077.} [\doug], Strominger
\REF\strom{A.~Strominger, 
``Open $p$-Branes'', preprint hep-th/9512059.}
[\strom], Townsend 
\REF\towns{P.~K.~Townsend, ``$D$-Branes from $M$-Branes'', preprint 
DAMTP-R-95-59, hep-th/9512062.} [\towns] and Vafa
\REF\vafa{C.~Vafa, ``Instantons on $D$-Branes'', preprint 
HUTP-95-A049, hep-th/ 9512078.} [\vafa]
showed that $p$-branes can have boundaries on $p$-branes.
More concretely, Strominger 
[\strom] and Townsend [\towns] 
suggested that the fivebrane may have an 
interpretation as a $D$-brane of an open membrane 
in eleven dimensions. 

One problem that we would like to address in this paper is to show
that the dynamics of the fivebrane can be 
obtained from the quantization of a ``small membrane''
after imposing appropriate boundary conditions on the 
fields. In section~2 we formulate boundary conditions 
for an open
membrane whose boundaries lie on the fivebrane of 
${\cal M}$-theory.
For this purpose we will use the light-cone gauged 
fixed action of the 
supermembrane. We will consider two particular 
classical configurations in
the next two sections. In section~3 we show that 
the massless fields on the 
worldvolume of the fivebrane can be obtained from 
the quantization of the 
fermionic zero-modes of a pointlike collapsed 
membrane that is confined to the fivebrane.
In section~4 we discuss the boundary dynamics of
 an infinitely extended planar membrane
that is stretched between two parallel fivebranes.
In appendix~A we use the Siegel gauge to show that 
the fermionic zero-modes
exhibit the full $SO(5,1)$ chirality of the worldvolume fields
of the fivebrane.
Our notation and conventions are explained 
in appendix~B. 
 
\chapter{Boundary Conditions for the Open Membrane}

\no The action for the eleven-dimensional supermembrane 
in flat superspace is
given by 
\REF\beseto{E.~Bergshoeff, E.~Sezgin and P.~K.~Townsend, 
``Supermembranes and Eleven-Dimensional Supergravity'', \pl 
{\bf B189} (1987) 75.}
\REF\annals{E.~Bergshoeff, E.~Sezgin and P.~K.~Townsend, 
``Properties of the Eleven-Dimensional Supermembrane Theory '', 
{\it Ann. Phys.} {\bf 185} 
(1988) 330.}[\beseto,\annals]: 
$$
\eqalign{
{\cal S}=-{1\o 2} \int
 d^3 \s  \sqrt{-h}\Bigl[  & h^{\a\b} \Pi_{\a}^M {\Pi_{\b}} _M  
-1+i \e^{\a\b\g} \Tbar \G_{MN} \pa_{\a} \T  \Bigl(
\Pi^M_{\b} \Pi_{\g}^N \cr 
& +i \Pi_{\b} ^M \Tbar\G^N \pa_{\g} \T 
-{1\o 3} \Tbar \G^M \pa_{\b} \T \Tbar \G^N \pa_{\g} \T
\Bigr)\Bigr], \cr}
\eqn\ai
$$ 
In the above expression $\T$ is a 32-component Majorana spinor, 
$X^{M}(\s)$, with $M=0,\dots,10$, describes the 
bosonic membrane configuration 
and $h_{\a\b}$, with $\a,\b=0,1,2$, 
is an auxiliary worldbrane metric with Minkowski signature. Furthermore
$$
\Pi^M_{\a}=\pa_{\a} X^M -i \Tbar \G^M \pa_{\a} \T. 
\eqn\aai
$$
Henceforth we neglect higher order terms in fermionic variables.

This action is invariant under worldvolume general coordinate 
transformations and under Poincar\'e transformations in the 
eleven-dimensional Minkowski space-time. Up to surface 
terms (whose cancellation depends on the boundary conditions chosen) 
it is also invariant under global space-time supersymmetries:
$$
\eqalign{
& \d _{\e}\T =\e, \cr
& \d_{\e}X^M= i {\bar \e} \G^M \T, \cr}
\eqn\bi
$$
as well as local $\k$-symmetries
$$
\eqalign{ 
& \d_{\k} \T =2P_+  \k(\s),  \cr
& \d_{\k} X^M =2i {\bar \T} \G^M P_+ \k(\s). \cr}
\eqn\bii
$$ 
Here $\e$ and $\k(\s)$ are eleven-dimensional Majorana spinors
that are worldvolume scalars. 
$P_{\pm}$ are projection operators 
\REF\hlp{J.~Hughes, J.~Liu and J.~Polchinski, ``Supermembranes'', 
\pl {\bf B180} (1986) 370.}
[\hlp] defined by 
$$
P_{\pm}= {1\o 2} \left(1\pm  {1 \o 3! }\e^ {\a\b\g} \pa_{\a} X^{M}
\pa_{\b} X^{N} \pa_{\g} X^{P} \G_{MNP}\right). 
\eqn\aiv
$$

The equations of motion are the embedding equation for the 
worldbrane metric 
\REF\duff{M.~J.~Duff, T.~Inami, C.~N.~Pope, E.~Sezgin and 
K.~S.~Stelle'', ``Semiclassical Quantization of the Supermembrane'', 
\np {\bf B297} (1988) 515.}
[\annals,\duff]: 
$$
h_{\a\b} = \pa_{\a} X^M \pa_{\b} X_M , 
\eqn\biii
$$ 
and the brane-wave equations
$$
\eqalign{
& \pa_{\a} \left( \sqrt{-h} h^{\a\b} \pa_{\b} X^M \right)=0, \cr
& P_- h^{\a\b} \pa_{\a} X^M \G_M \pa_{\b} \T=0.  \cr}
\eqn\ci
$$
Since $P_-$ is a projection operator,
the second equation in {\ci} is an equation of motion 
for only 16 of the 
32 components of $\T$. This is due to the fact 
that the action has a $\k$-symmetry. The local invariances 
of the action can be used to impose  
the string inspired light-cone gauge 
\REF\hoppe{J.~Hoppe, ``Quantum Theory of a Relativistic Surface'', 
Aachen preprint PITHA 86/24 and Ph.D. thesis, MIT 
(1982).} [\hoppe,\annals]:
$$
X^+ =p^+ \t, \qquad 
h_{\a\b} =\left( \matrix{-\det g & 0\cr 0 & g_{ab}\cr}\right)
\qquad {\rm and} \qquad \G^+\T=0,
\eqn\gplu
$$
where $a,b=1,2$. Choosing an adequate representation of the gamma 
matrices the field $\T$ can be written in the form $\T=(0,S)$, 
where $S$ is a real 16-component $SO(9)$ spinor. 
Thus half of the 32 components of 
$\T$ have been eliminated. We can further eliminate three 
bosonic degrees of freedom. The first  
equation in {\gplu} eliminates $X^+$ as independent coordinate 
and inserting the gauge conditions into the embedding equation 
one finds that $X^-$ satisfies 
\REF\bapose{I.~Bars, C.~N.~Pope and E.~Sezgin, ``Massless Spectrum 
and Critical Dimension of the Supermembrane'', \pl 
{\bf B198} (1987) 455.}
[\bapose,\annals]:
$$
\eqalign{ 
{\dot X^-} & ={1\o 2p_+} \left( {\dot X^I } {\dot X_I} + \det g
\right),  \cr
\pa_a X^- & ={1\o p_+} {\dot X^I} \pa_a X_I ,  \cr}
\eqn\qi
$$
where $I=1,\dots,9$. This determines $X^-$ and furthermore  
from the second equation one obtains the constraint 
$$
\e^{ab} \pa_a {\dot X}^I \pa_b X_I 
=0, 
\eqn\qii
$$
which can be used to eliminate an additional coordinate.
In the following we will take this variable to be $X^1$
without loss of generality. Therefore 
we are left with eight bosonic degrees of freedom. In the 
light-cone gauge the covariant field equations {\biii} and {\ci} 
take the form
$$
\eqalign{
& g_{ab} =\pa_a X^I \pa _b X_I,  \cr
& {\ddot X^I} =\pa_a (g g^{ab} \pa_b X^I), \cr
&{\dot S} =-\e^{ab} \pa_a X^I \g_I \pa_b S. \cr}
\eqn\qiii
$$
The light-cone gauge fixed action has residual symmetries, that 
leave the gauge conditions {\gplu} and the 
equations of motion {\qiii}
invariant. The form of these transformations has been obtained in 
[\annals]. They are the $\a$-symmetries:
$$
\eqalign{
& \d S =-{1\o \sqrt{2} p^+} \left( {\dot X}^I \g_I 
-{1\o 2} \e^{ab} \pa_a X^I \pa_b X^J \g_{IJ} \right) \a ,\cr
& \d_{\a} X^I=2i {\bar \a} \g^I S+2i\e^{ab} \pa_a X^I 
{\bar \a}\int_0 ^{\tau} d \t \pa_b S,\cr}
\eqn\qiv
$$
and the $\b$-symmetries:
$$
\eqalign{
& \d_{\b} S=\b,\cr
&\d_{\b} X^I=0.\cr}
\eqn\qv
$$
Here $\a$ and $\b$ are 16-component spinors resulting 
from the decomposition $\e =(i\a ,\b )$. 
From the above symmetries only the first one 
is a supersymmetry.

Consider now the variation of the action under a variation 
of the bosonic fields.
Demanding the equation of motion for $X^I$ to hold, the variational 
principle for the action is:
$$
\int\nolimits_{\pa \S} n^a \pa_a X_I \d X^I=0, 
\eqn\axi
$$
where $n^a$ is the outward-pointing unit normal 
to the boundary. We will take $n^a$ to be spacelike everywhere. 
This imposes constraints for the values 
of the fields at the boundary. 
For the bosonic fields we could in principle impose either 
Neumann boundary 
conditions, where we allow $\d X^I$ to be arbitrary and the 
normal derivative 
vanishes, or Dirichlet boundary conditions where $X^I$ is 
held fixed at the boundary.
However, it was shown in [\strom] that RR charge conservation
does not allow the existence of free open membranes 
\foot{An exception of this is
the open membrane considered in [\howi], where the 
boundaries lie at the end of space-time.}.  
Charge conservation allows a membrane ending on one 
fivebrane. When the fivebrane lies in the hyperplane 
$X^M=0$ for $M=1,\dots,5$, the boundary conditions for the physical 
bosonic fields are
$$
\eqalign{
\XD & =0 , \qquad I =2,3,4,5, \cr 
n^a \pa_a \XN & =0, \qquad I =6,7,8,9,
\qquad {\rm on} \;\; \pa\S .\cr } 
\eqn\axii
$$

Next we would like to determine the 
boundary conditions for $S$\foot{Boundary conditions for 
the fermionic fields of an open spinning membrane were discussed 
previously in 
\REF\lumo{H.~Luckock and I.~Moss, ``The Quantum Geometry 
of Random Surfaces and Spinning Membranes'', {\it Class. Quantum 
Grav.} {\bf 6} (1989) 1993.}[\lumo].}. 
The vanishing of $\d {\cal S}$ under the variation of the fermionic 
field requires
$$
\int\nolimits_{\pa \S} {\bar S} \tsl \d S=0, 
\eqn\axiii
$$
where $\tsl=t^a \pa_a  \XN \g_I$. Since the Dirac operator 
is first order, we are only allowed to fix half of the 
components of the fermionic field at the boundary.
For this purpose we
introduce boundary projection operators
$$
\wp_{\pm}={1\o 2} ( 1\pm \g_{6789} ) , 
\eqn\axiv
$$
which act on spinor fields at the boundary.  
These operators satisfy 
$$
\wp_{\pm}^2 =\wp_{\pm} , \qquad 
\wp_+ \wp_- =0 \qquad {\rm and} \qquad 
 \wp_++\wp_-=1.
\eqn\di
$$
The expression {\axiii} vanishes if we impose the Dirichlet
condition
$$
\wp_+ S =0 \qquad {\rm on }\;\; \pa \S. 
\eqn\me
$$
This can be easily seen using {\axiv} and the ``flipping'' property  
of appendix B. From {\qiv} we see that 
the supersymmetry generating
parameter satisfies the boundary condition
$$
\wp_- \a =0 \qquad {\rm on }\;\; \pa \S. 
\eqn\mei
$$
This implies that half of the supersymmetries are broken in the 
presence of the boundary, as it happens for the open superstring.
The above equations state that the fermionic fields have a well 
defined $SO(4)$ chirality, longitudinal to the fivebrane.
This choice of fermionic boundary conditions guarantees that 
on the boundary $\d _{\a} \XD=0$, while there is no restriction 
on the Neumann coordinates. 
This means that the residual supersymmetry 
transformation respects 
the Dirichlet boundary conditions on the bosonic fields, 
as it has to be.

To show that the Neumann boundary conditions are compatible 
with the residual supersymmetries, we first have to find the
boundary conditions for the orthogonal component 
of the fermionic field. 
They can be obtained using the Dirac equation  
and the boundary condition {\me}. Here we have 
to extend the definition of
$n^a$ and $t^a$
to a neighborhood of $\pa \S$ by parallel 
transport along geodesics normal to
$\pa \S$, so that $n^a \nabla_a n^b =n^a \nabla_a t^b$. 
Then, it is easiest to 
start with the expression 
$$
\wp_+\left({\dot S}+\e^{ab} \pa_a X^I \g_I \pa_b S\right)=0, 
\eqn\fvi
$$
that can be further transformed taking into account that the derivative 
of {\me} with respect to the tangent along the boundary is equal 
to zero. This leads us to 
$$
n^a \pa_a (\wp_- S)=0\qquad {\rm on} \;\; \pa\S . 
\eqn\fv
$$
This boundary condition implies that $n^a \pa_a( \d_{\a} \XN)=0$
and is therefore compatible with the bosonic Neumann 
boundary condition in {\axii}.

Notice that to derive the boundary conditions on the
 bosonic and fermionic fields we did not need 
to specify a particular classical configuration.
The specific form of the classical configuration, 
together with the boundary condition {\me} 
determines the number of unbroken supersymmetries 
for a particular
brane configuration. In the following we will consider 
two particular 
classical membrane configurations: a collapsed membrane or 
zero-brane that is
confined to a single fivebrane and an infinitely extended 
planar membrane that is
stretched between two parallel fivebranes.   
 
\chapter{The Zerobrane Confined to the Fivebrane}

\no Some time ago, Bars, Pope and Sezgin [\bapose]
obtained the massless spectrum of eleven-dimensional 
supergravity from the 
quantization of a closed membrane that was completely 
collapsed to a point.
Here we would like to show that the 
massless fields on the fivebrane worldvolume 
can be obtained from the quantization of a zerobrane which is
confined to the fivebrane and that is  
traveling with the 
speed of light.
This collapsed membrane can be viewed as the groundstate of an open
membrane having boundaries on the fivebrane. 
\foot{We believe that the topology of the extended membrane
 whose groundstate 
is described by this zerobrane is a cylinder in
$\IR^{10} \times S^1$ which has one end lying on the fivebrane 
and is closed at the other end. However, further analysis is
required to confirm this. This would be relevant 
to understand the massive spectrum of
the theory.}

In the limit where the membrane is collapsed to 
a point
the bosonic part of the classical field configuration 
takes the form
  
$$
X^M(\t,\s_1,\s_2)=x^M+p^M \t, 
\eqn\qiv
$$
where $x^M$ and $p^M$ are constants. For this configuration 
the metric becomes degenerate, but nevertheless 
the field equations are well defined, so that this 
solution is perfectly regular. 
The equations of motion {\biii} and {\ci} in the light-cone gauge, 
reduce 
to the equations of motion of a massless superparticle 
\REF\gsw{M.~Green, J.~H.~Schwarz and E.~Witten, ``Superstring Theory'', 
Cambridge Univ. Pr. (1987), Volume I.}[\bapose,\gsw], 
$$
p^2=0, \qquad {\dot p}^M =0 \quad {\rm and} \quad p_M \G^M
{\dot \T}=0,  
\eqn\qv
$$
which is moving with the speed of light. 
Due to this fact, the spectrum of the supermembrane contains massless 
particles.
The Dirichlet conditions 
on the bosonic coordinates {\axii} translate to the 
constraint
$$
 X^M\equiv 0 \quad {\rm for} \quad M=2,3,4,5,
\eqn\qvvii
$$
which means that 
the particle is confined to the fivebrane. The residual 
supersymmetries 
for the configuration {\qiv} take the form: 
$$
\eqalign{
&\d_{\a}  S   =-{1\o \sqrt{2} p^+} {\dot X} ^I \g_I\a, \cr
& \d_{\a}  X^I =2i{\bar \a} \g^I S. \cr }
\eqn\qqv
$$
Using this $\a$-symmetry and the bosonic constraint {\qvvii} we obtain
the conditions
$$
(1+\g_{6789}) S=(1-\g_{6789})\a=0. 
\eqn\qqvi
$$ 
Thus the fermionic fields have a well defined $SO(4)$ 
chirality longitudinal 
to the fivebrane. 

The massless spectrum of the theory can be determined from the 
quantization of the fermionic zero-modes obtained from {\qqv}. 
In the 
light-cone gauge there are 16 real fermionic zero-modes 
that are subjected to the 
constraint {\qqvi}. Therefore only 
eight linearly independent fermionic zero-modes are left. 
Since the light-cone gauge 
fixed action contains the term ${\bar S} {\dot S}$, 
the canonical conjugate to real $S$ 
is again $S$, so that the zero-modes satisfy a Clifford algebra 
$$
\{ S^{\a}_0 , {S_0}_{\b} \}=2\d^{\a}_{\b}. 
\eqn\axix
$$ 
These fermionic zero-modes can be rearranged into four creation and
four annihilation operators. The corresponding 
Hilbert space will 
be $2^4$-dimensional with $2^3$ bosons and $2^3$ fermions. 

We would like to argue that the zero-modes {\axix} 
subjected to the constraint 
{\qqvi} generate a chiral $N=2$, $d=6$ spectrum corresponding to 
the collective coordinates on the fivebrane worldvolume. 
There are two $N=2$ theories in $d=6$ depending on their field 
content 
\REF\hosito{P.~S.~Howe, G.~Sierra and P.~K.~Townsend, 
``Supersymmetry
in Six-Dimensions'', \np {\bf B221} (1983) 331.}
\REF\ro{L.~J.~Romans, ``Selfduality for Interacting Fields: 
Covariant Field Equations
for Six-Dimensional Chiral Supergravity'', \np {\bf B276} 
(1986) 71. }
\REF\world{C.~G.~Callan, J.~A.~Harvey and A.~Strominger, ``Worldbrane 
Actions for String Solitons'', \np {\bf B367} (1991) 60.}
\REF\gito{G.~W.~Gibbons and P.~K.~Townsend, ``Vacuum Interpolation 
in Supergravity Via Super $p$-Branes'', \prl {\bf 71} (1993) 3754.}
\REF\kami{D.~Kaplan and J.~Michelson, ``Zero-modes for the $D=11$ 
Membrane and Fivebrane'', preprint hep-th/9510053.}
[\hosito,\ro,\world,\gito,\kami]. There are three different matter 
multiplets, $\Phi$, $A$ and $B$ [\world]. Each 
of these matter multiplets contains a doublet of Weyl fermions. 
In the $(1,1)$ theory these fermions have different chirality 
while they have the same for the $(2,0)$ theory. $\Phi$ contains 
four scalars, $A$ a single vector and $B$ contains an anti-self-dual 
tensor field and a single scalar. 

The $(0,2)$ theory has a matter 
multiplet which is the sum of the $\Phi$ and 
$B$ multiplet. The anti-self-dual tensor has 
three degrees of freedom 
which together with the five scalars matches the eight 
bosonic degrees 
of freedom that we get out of the representation of the 
Clifford algebra. 
The constraint {\qqvi} indicates that the fermionic zero-modes 
have a well defined chirality longitudinal to the fivebrane. 

Therefore we have generated 
the chiral $(0,2)$ $N=2$ theory in the light-cone gauge, 
corresponding to the massless
states on the six-dimensional 
worldvolume of the fivebrane! 
This shows that the fivebrane has indeed an interpretation as
a $D$-brane of an open membrane in eleven dimensions.

\chapter{The Membrane Stretched Between Parallel Fivebranes}
\no In addition to a membrane with boundaries on a 
single fivebrane, we could 
consider a membrane that is stretched between two 
(or more) parallel fivebranes.
An infinitely extended planar membrane that is 
stretched between e.g. two parallel
fivebranes corresponds to a BPS state that 
preserves one quarter of the original
supersymmetries to leading order [\strom].  
It was argued in [\strom,\towns] that the 
boundary dynamics of the open membrane is 
described by a six-dimensional non-perturbative 
superstring theory 
\REF\dfk{M.~J.~Duff, S.~Ferrara, R.~Khuri and 
J.~Rahmfeld ``Supersymmetry and Dual String 
Solitons'', \pl {\bf B356} 
(1995) 479, hep-th/9506057.}
[\dfk].
This string becomes tensionless as the two 
fivebranes approach each other. 
The boundary conditions for the bosonic fields 
of this open membrane are described
by two equations of the type {\axii}, one for 
each boundary. The fermionic fields 
obey the boundary conditions {\me} and {\fv}.
Thus half of the original field $\T$ propagates on the fivebrane
and that half has a definite $SO(4)$ chirality in light-cone gauge. 
Upon double dimensional 
reduction to ten dimensions along the fivebranes,
we are left with a
string stretched between two Dirichlet four-branes
of the Type IIA string theory.

\vskip 0.2 cm
\no {\bf Acknowledgments} 

\no We are grateful to I.~Bars, M.~Douglas, M.~Duff, 
J.~Gauntlett, G.~Moore, 
J.~Polchinski and specially to 
A.~Strominger for useful discussions. This work was 
supported 
by DOE grant DOE-91ER40618 and NSF grant 
PHY89-04035. 

\endpage
\no {\bf Appendix A: Boundary Conditions in Siegel Gauge}

\no For completeness we want to show how to get the full $SO(5,1)$ 
chirality for an open extended membrane. For this purpose it will 
be convenient to work in the so-called Siegel gauge for the 
fermionic field [\annals]. The $\k$-symmetry can be used to set 
$$
P_+ \T=0. 
\eqn\Bi
$$
We are then left with 16 physical components for $\T$.
Now we want to determine the boundary conditions for $\T$. 
The vanishing of $\d {\cal S}$ under the variation 
of the fermionic 
field requires
$$
\int\nolimits_{\pa \S} {\overline \T} \nsl \d \T=0, 
\eqn\Bii
$$
where $\nsl=n^{\a} \pa_{\a}  X_{\SSD}^M \G_M$.
The expression {\Bii} vanishes if we impose the Dirichlet
condition
$$
(1+ \prod_{M\in {\cal N}} \G_M ) 
\T=0 \qquad {\rm on }\;\; \pa \S. 
\eqn\Biii
$$
This equation implies that the fermionic fields have a well 
defined $SO(5,1)$ chirality, longitudinal to the fivebrane. 

The boundary conditions for the orthogonal component 
of the fermionic field can be obtained from the vanishing of the 
boundary term
$$
\int \nolimits_{\pa \S} \d X_{\SSN}^M
\Tbar \G_M n^{\a} \pa_{\a}  \T =0. 
\eqn\Biv
$$
which leads us to 
$$
n^{\a} \pa_{\a} \left( \wp_- \T \right) =0 \qquad {\rm on} 
\;\; \pa \S. 
\eqn\Bv
$$
This boundary condition can be obtained out of the Dirac 
equation and the boundary condition {\Biii}.

\endpage
{\bf Appendix B: Notation and Useful Formulas} 
\item{\tria}The space-time gamma matrices satisfy the Clifford 
algebra 
$$\{ \G^M,\G^N\}=2\eta ^{MN}, 
$$
with signature $(-,+\dots,+)$. 
$\G^0$ is antihermitian and the rest of the gamma matrices 
are hermitian with
$$
{\G^M}^{\dagger} =\G^0 \G^M \G^0
$$
and for an arbitrary spinor we define ${\bar \Psi}=\Psi^{\dagger} 
\G^0$. We use the notation
$$
\G^{M_1\dots M_n}=\G^{[M_1}\G^{M_2} \dots \G^{M_n]},
$$
where the square bracket implies a sum over
$n!$ terms with an $1/n!$ prefactor.
\item{\tria}In the light-cone gauge we use the notation
$$
X^{\pm}={1\o \sqrt{2}} \left(X^0 \pm X^{10}\right)  \qquad {\rm and} 
\qquad  
\G^{\pm}={1\o \sqrt{2}}\left(\G^0\pm \G^{10}\right), 
$$
with $(\G^{\pm})^2=0$, $\{\G^+,\G^-\}=-2$. 
The gamma matrices are decomposed according to 
$$
\G^+=I_{16} 
\otimes \left( \matrix{ 0 & 0 \cr \sqrt{2} i & 0 \cr }\right), 
\quad 
\G^-=I_{16} \otimes \left( \matrix{ 0 & \sqrt{2} i  \cr 0 & 0 \cr }
\right), 
 \quad  
\G^I=\g^I\otimes \left( \matrix{1 & 0 \cr 0& -1 \cr} \right) 
$$
and they satisfy $\{\g^I,\g^J\}=2\d^{IJ}$. Set 
$\T=(i\T_1, \T_2)$ and 
${\bar \T} =(-i {\bar \T}_2, -{\bar \T}_1)$. We use the ``flip'' 
rule:
$$
{\bar \th} \g^{\n_1 \dots \n_n}\psi =(-)^{{n(n-1)\o 2}+1 }
{\bar \psi} \g^{\n_1 \dots \n_n} \th . 
$$
\item{\tria}A useful identity is
$$
\G \G_M \pa^{\g} X^M=\pa^{\g} X^M\G_M \G  
={1\o 2} \e^{\a\b\g}\pa_{\a}X^R \pa_{\b} X^S 
\G_{RS}. 
$$

\endpage
\refout
\end